\documentclass[10pt,a4paper,final]{iopart}
\usepackage{iopams}

\expandafter\let\csname equation*\endcsname\relax

\expandafter\let\csname endequation*\endcsname\relax

\usepackage{mathrsfs,amsmath}
\usepackage[scr=boondox]{mathalfa}
\usepackage{cite,comment}
\usepackage{hyperref}
\usepackage{graphicx, float}% Include figure files and allows positioning
\usepackage{dcolumn}% Align table columns on decimal point
\usepackage{bm}% bold math
\usepackage{makecell}
\usepackage{epstopdf}
\usepackage[usenames, dvipsnames]{color} % Change the colour of elements
\usepackage[table, dvipsnames]{xcolor} %for use in color links
\usepackage{colortbl}
\usepackage[caption=false]{subfig}
\usepackage{booktabs}

\makeatletter
\let\@fnsymbol\@arabic %Set footnotes to arabic numerals
\renewcommand{\@makefnmark}{\hbox{\textsuperscript{\tiny{\@thefnmark}}}} %Reducing the size of the footnote mark
\makeatother

\renewcommand{\footnoterule}{% Produces the rule separating the main text on a page from the page’s footnotes
	\kern -3pt                         % This -3 is negative
	\hrule width \textwidth height 1pt % of the sum of this 1
	\kern 2pt}                         % and this 2

\makeatletter
\newcommand{\hwidth}[1]{%
	\noalign{\hrule \@height #1}%
}
\makeatother

\begin{document}
\title[The TOV equations in isotropic coordinates]{\large{The representation and computational efficiency of the Tolman--Oppenheimer--Volkoff equations in isotropic coordinates}}

\author{D\'{a}niel Barta}
\address{HUN-REN Wigner RCP, H-1525 Budapest 114, P.O. Box 49, Hungary}
\ead{barta.daniel@wigner.hun-ren.hu}
\vspace{10pt}

\begin{abstract}
This study aims to provide an analytical scheme for computing equilibrium configurations of relativistic stars by solving the Tolman--Oppenheimer--Volkoff equations directly in isotropic polar coordinates, as opposed to the commonly applied methods of rescaling the radial profile of corresponding solutions obtained in curvature coordinates. This study also provides evidence that the differential equation for gravitational mass may be replaced by an algebraic expression relating the metric potential to the energy density in the form of the quartic equation. Nevertheless, the greater computational expense of evaluating the algebraic equation renders its application less efficient. A further objective of this study was to evaluate the performance of the present computational scheme in the computational time and numerical accuracy. Our results indicate that the computational time increases with the stiffness of the constituent matter inside the star. Conversely, the absolute difference between the gravitational mass obtained by the proposed method and that computed via the use of LORENE packages initially increases rapidly with the central energy density, but the rate of growth subsequently declines as the maximum mass configuration is approached.
\end{abstract}

\section{\label{sec:intro}Introduction}

Over a century has passed since the Einstein field equations (EFEs) for a perfect-fluid body in a static, spherically symmetric spacetime were first formulated in isotropic coordinates by Arthur Eddington in 1923. \cite{Eddington1923} This formulation of the EFEs, which consists of a coupled system of second-order ordinary differential equations (ODE) in the metric variables, also appears in Richard Tolman's two research papers \cite{Tolman1930a,Tolman1930b} in 1930, as well as Georges Lema\^{i}tre's famous monograph \cite{Lemaitre1933}, where he recognised in 1931 that the mass function is interchangeable by the coefficient of the spatial metric function. It has been realised by these early works \cite{Tolman1930a,Tolman1930b,Lemaitre1933}\nocite{Tolman1939,Oppenheimer1939} that two of the three non-zero components of the EFEs are second-order ODEs in the metric variables and, according to Ref. \cite{Herrera2016}, can regarded merely as a rather elementary extension of energy--momentum conservation. Consequently, albeit one of the ODEs is replaced by the equation of local conservation of energy--momentum, there is still at least one second-order equation left, making exact solutions either analytically or numerically unattainable. \cite{Semiz2017} The issue of finding exact solutions for these EFEs was first addressed by Tolman \cite{Tolman1939}, and independently by J. Robert Oppenheimer and George Volkoff \cite{Oppenheimer1939} in 1939. The latter revealed that both the use of curvature coordinates (also called Schwarzschild coordinates) in the EFEs and introducing the function of gravitational mass considerably simplify the ODE system \cite{Semiz2016} to a system of two first-order differential and one algebraic equations, with the boundary condition of all three variables.\footnote{Let us note that a recent correspondence \cite{Semiz2016,Herrera2016,Semiz2017} between Semiz and Herrera argued that instead of ``Tolman--Oppenheimer--Volkoff equation,'' the ``Oppenheimer--Volkoff equation'' be a more suitable name for the solutions to the problem at hand, but this study does not wish to take side over the appropriate nomenclature.}

An extensive list of exact solutions to the EFEs had been collected in the literature (see e.g. \cite{Delgaty1988,Stephani2003} and references therein), where the solutions are related to some specific restrictions imposed on the geometry of the spacetime (such as spherical or axial symmetries) or a particular distribution of matter (such as an isotropic distribution) within the spacetime. However, not all of them can be considered as physically acceptable solutions, only those that satisfy a certain set of conditions that serve to maintain the causality of the spacetime structure and the energy conditions. The number of suitable solutions can be further reduced by considering only those exact solutions that are associated with isolated spheres of perfect fluid with boundary conditions imposed at a finite radius. Even though some of these exact solutions describe identical geometries in different coordinate frames, the transformation from one frame to another, as argued by Ref. \cite{LasHeras2019}, is not always well-defined.\footnote{Specifically, Ref. \cite{Nariai1950} emphasised that Schwarzschild-like coordinates associated to an arbitrary line element can always be transformed into isotropic polar coordinates, but the reverse process cannot always be performed.} The first exact solutions in isotropic coordinates was discovered in 1950 by Nariai \cite{Nariai1950}, where one is identical with Tolman's solution III\footnote{This solution is the Schwarzschild interior solution, which also includes Tolman's solution I (also known as Einstein's static universe), as a special case where the constant of integration $c$ is set equal to zero.}, another one is associated with new static and homogenous cosmological solution, and most importantly, one represents an entirely new type of ``stellar'' solution involving trigonometrical and logarithmic functions of the radial coordinate. Further regular and well-behaving exact solutions in isotropic coordinates were found by Refs. \cite{Goldman1978} for spherically symmetric distributions of isotropic perfect fluid.

Despite the fact that the TOV equations in the literature are most commonly expressed in terms of curvature coordinates, the quasi-isotropic polar coordinates are a more commonly preferred choice to initiate dynamical calculations that are implemented in a formalism based on the $3+1$ decomposition. The efficiency of numerical methods in these codes is greatly improved by being performed in quasi-isotropic coordinates, where the three spatial dimensions are treated similarly. One of the well-known examples is the \href{https://lorene.obspm.fr}{\texttt{LORENE}} code \cite{LUTH1997} which uses a multi-domain spectral method to solve the EFEs decomposed to a system of five quasi-linear elliptic equations. Ref. \cite{Bonazzola1993} established that one may adapt maximal-slicing quasi-isotropic (MSQI) coordinates in a stationary and circular axisymmetric spacetime for the description of the 3+1 setting of EFEs. As a result, the solution to the EFEs is directly obtained in quasi-isotropic coordinates but not through solving the set of equations corresponding to the TOV equations. Another core computational tool in relativistic astrophysics is the \href{https://einsteintoolkit.org/}{\texttt{Einstein Toolkit}} \cite{EinsteinToolkit:2023_05}, which applies the modular \texttt{Cactus} framework, consisting of general modules called ``thorns.'' The \texttt{TOVSolver} thorn solves the standard TOV equations expressed in curvature coordinates, then transforms every variable into isotropic coordinates by integrating the radius conversion formula $\displaystyle \frac{\partial \log(r/\tilde{r})}{\partial\tilde{r}} =  \frac{\tilde{r}^{1/2}-(\tilde{r}-2m)^{1/2}}{\tilde{r}(\tilde{r}-2m)^{1/2}}$, which is subject to the boundary condition \eqref{radius-transformation-exterior} in the exterior region of spacetime. \cite{Loffler2012} In converting the solution into the variables required for a dynamical evolution, one may assume that the metric is conformally flat, with a conformal factor given by $\psi = \tilde{r}/r$, or equivalently, a logarithmic conformal factor $\phi = \frac{1}{2}\log(\tilde{r}/r)$. The \texttt{Einstein Toolkit} framework contains three routines that can read in publicly available data generated by the \texttt{LORENE} code. Although \texttt{LORENE} was originally developed for computing rapidly rotating neutron stars, it has been gradually extended to be capable of computing various kinds of equilibrium and quasi-equilibrium initial data sets of single or binary compact objects. Another code, \texttt{COCAL} was developed by Ury\={u} \& Tsokaros \cite{Uryu2012} for the very purpose of computing such initial data sets. The numerical method used in \texttt{COCAL} for computing equilibrium configurations is based on an extension of the Komatsu--Eriguchi--Hachisu (KEH) method \cite{Komatsu1989a,Komatsu1989b} which is less technical than the spectral methods utilised in \texttt{LORENE}. Ref. \cite{Tsokaros2015} explains that instead of using spectral methods in a 3+1 setting, the one-dimensional KEH solver seeks solutions for the TOV equations in terms of curvature coordinates, then rescales the radial profile of the solution to that of an isotropic setting by $d\tilde{r}/dr = (\tilde{r}/r)\sqrt{1-2m(\tilde{r})/\tilde{r}}$. The Appendix B of Ref. \cite{Tsokaros2015} briefly explains the transformation of the TOV equations from curvature coordinates to isotropic coordinates, which we have reproduced using our notations for the sake of comparison in the discussion of our results in Sec. \ref{sec:conclusion}.

The present study aims, on the one hand, to present a partially uncoupled system of first-order ODEs, derived in isotropic coordinates, which corresponds to the TOV equations written in curvature coordinates. This involves proving that the differential equation for gravitational mass may be replaced by an algebraic expression relating the metric potential to the energy density, even in the context of isotropic coordinates. On the other hand, this study is intended to outline certain desirable attributes of computing equilibrium configurations by directly solving the TOV equations in isotropic coordinates rather than applying a method of rescaling the radial profile of corresponding solutions obtained in curvature coordinates.

\section{\label{sec:coordinate-frames}Centrally symmetric spacetime represented in different types of coordinate frames}
Let us consider a gravitational field that possesses central symmetry. This aspect is expressed through the square of the line element, which is given by
\begin{equation} \label{line-element-definition}
	ds^{2} = A(r)dt^{2} + B(r)dr^{2} + 2C(r)dr dt + D(r)d\Omega^{2},
\end{equation}
where $A,\, B,\, C,\, D$ are all certain functions of the radial coordinate $r$ and
\begin{equation} \label{two-sphere-metric}
	d\Omega^{2} = \sin^2\theta d\phi^{2} + d\theta^{2}
\end{equation}
denotes the volume element for the 2-sphere of unit radius. One may subject the coordinates to any transformation that does not destroy the central symmetry of \eqref{line-element-definition} over the spatial coordinates. This implies that a transformation $r = r(\tilde{r})$ can be found where $r$ is any function of the new coordinate $\tilde{r}$.\footnote{See the transformation laws \eqref{radius-transformation-interior2} and \eqref{radius-transformation-exterior} for the interior perfect-fluid region and for the exterior vacuum region of the Schwarzschild spacetime, respectively.} One may choose the radial coordinate in such a way that:
\begin{enumerate}
	\item the metric function $C(\tilde{r})$, being the coefficient of the term $2drdt$, vanishes and the coefficient $D(\tilde{r})$ becomes equal to $\tilde{r}^{2}$;
	\item or the metric function $C(r)$ vanishes once again, but this time the metric function $D(r)$ will be identical to $B(r)$.
\end{enumerate}

Having the remaining metric-potential functions conveniently expressed in exponential forms as 
\begin{equation} \label{exponential-form}
	A(\tilde{r}) = e^{2\tilde{\nu}(\tilde{r})},\, B(\tilde{r}) = e^{2\tilde{\lambda}(\tilde{r})} \quad \text{or} \quad A(r) = e^{2\nu(r)},\, B(r) = e^{2\mu(r)},
\end{equation}
respectively, the metric of an arbitrary static spherically symmetric spacetime can thus be expressed by spherical polar coordinates (called Schwarzschild coordinates) $(\tilde{t},\, \tilde{r},\, \tilde{\theta},\, \tilde{\phi})$ as 
\begin{equation} \label{spherical-line-element}
	ds^{2} = -e^{2\tilde{\nu}}d\tilde{t}^{2} + e^{2\tilde{\lambda}}d\tilde{r}^{2} + \tilde{r}^2(\sin^2\tilde{\theta} d\tilde{\phi}^{2} + d\tilde{\theta}^{2}),
\end{equation}
or equivalently, by isotropic polar coordinates $(t,\, r,\, \theta,\, \phi)$ as 
\begin{equation} \label{isotropic-line-element}
	ds^{2} = -e^{2\nu}dt^{2} + e^{2\mu}[dr^{2} + r^2(\sin^2\theta d\phi^{2} + d\theta^{2})].
\end{equation}
Setting side-by-side the line elements (\ref{spherical-line-element}--\ref{isotropic-line-element}), and by requiring that the angular and the radial parts to be equal, respectively, one obtains
\begin{equation} \label{radius-transformation-interior}
	\tilde{r}^{2} \equiv e^{2\mu}r^{2}, \quad e^{2\mu}dr^{2} \equiv e^{2\tilde{\lambda}}d\tilde{r}^{2}.
\end{equation}
The earlier identity in eqs. \eqref{radius-transformation-interior}, which stems from the equality of the angular parts, defines a relation
\begin{equation} \label{radius-transformation-interior2}
	r = \exp\left(\int e^{\tilde{\lambda}}\tilde{r}^{-1}d\tilde{r}\right) = \exp\left[-\operatorname{Ei}\left(-\tilde{\lambda}\tilde{r}\right)\right]
\end{equation}
between the radial coordinates of the two coordinate frames. In addition, the earlier identity in eqs. \eqref{radius-transformation-interior} implies another relationship between the derivatives:
\begin{equation} \label{radial-coordinate-derivative}
	d\tilde{r} = e^{\mu}(r\mu' + 1)dr.
\end{equation}
The later identity in eqs. \eqref{radius-transformation-interior}, with the substitution of $dr$ from \eqref{radial-coordinate-derivative}, yields 
\begin{equation} \label{lambda-correspondence}
	e^{2\tilde{\lambda}} = (r\mu' + 1)^{-2}.
\end{equation}
\begin{figure}
	\centering
	\includegraphics[scale=0.65]{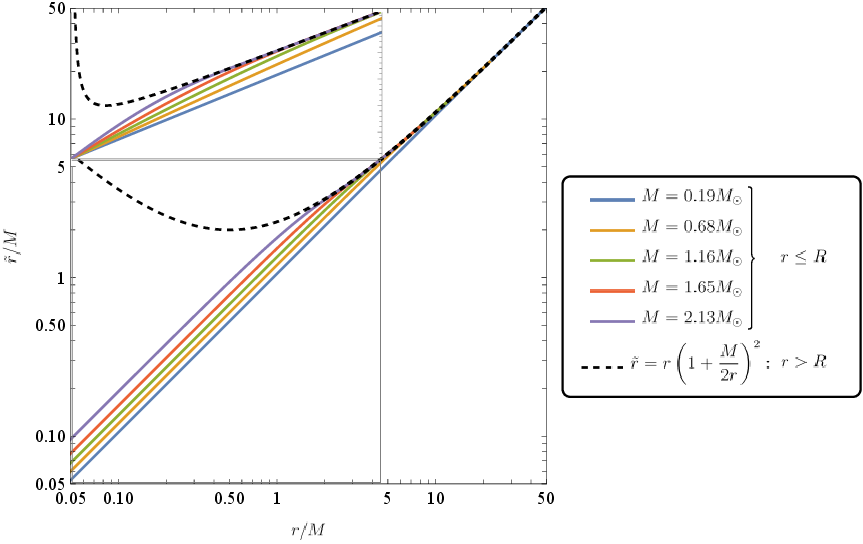}% Here is how to import EPS art
	\caption{The isotropic radial coordinate $r$ as a function of the curvature radial coordinate $\tilde{r}$ shown on logarithmic scale. In the interior region of the spacetime, the relation \eqref{radius-transformation-interior2} holds. This relation is displayed by a set of coloured lines; each line corresponds to a configuration of relativistic stars with different values of gravitational mass, and each line is marked by the bullet point symbol $\bullet$ in Figure \ref{fig:mass-radius-relation}, respectively. In the exterior region, the relation is simply expressed by eq. \eqref{radius-transformation-exterior} and represented by the dashed black line. The inset figure on top shows the the inner region $0.05 \leq r/M \leq 4.50$ on linear scale.}
	\label{fig:transformation}
\end{figure}
Fig. \ref{fig:transformation} shows the relationship between the radial coordinates of the two coordinate frames \eqref{radius-transformation-interior2} for five equilibrium stellar configurations constructed with the EoS SFHo/DD2, each with different values of gravitational mass. The figure implies that $\tilde{r}$ is very close to being a linear function of $r$ with an intercept of zero and a slope barely higher than one. The slope of line is initially larger for larger gravitational mass and it decreases slightly upon approaching the exterior region where it smoothly merges into the parabola \eqref{radius-transformation-exterior} (depicted by the dashed black line). The defining difference between the two choices of coordinate frames lies in the interpretation of the radial coordinate. The radial coordinate $\tilde{r}$ in \eqref{spherical-line-element} possesses a natural geometric property in terms of the surface $\tilde{A}$ by being conveniently defined as $\tilde{r} = \sqrt{\tilde{A}/4\pi}$. In contrast to that, the radial coordinate $r$ in eq. \eqref{isotropic-line-element} is defined so that the light cones on constant time-slices (or hypersurfaces) $\Sigma_{t}$ with $t = t_{0}$ appear as round spheres. The line element \eqref{isotropic-line-element} implies that the metric restricted to any of these hypersurfaces is 
\begin{equation} \label{nested-sphere}
	g|_{t = t_{0},\, r = r_{0}} = e^{2\mu(r_{0})}r_{0}^{2}g_{\Omega},
\end{equation}
where $g_{\Omega} \equiv d\Omega^{2}$ is the Riemannian metric on the 2-sphere of unit radius. This conveys that these nested coordinate-spheres \eqref{nested-sphere} do in fact represent geometric spheres, but the radial coordinate $r$ does not faithfully represent radial distances within the nested coordinate-spheres. On the other hand, angles in the constant time-slices (or hypersurfaces) are represented without distortion by the angular coordinates $(\theta,\,\phi)$ in isotropic charts. \cite{Gourgoulhon2007} This advantageous feature of (quasi-)isotropic coordinates is exploited in the 3+1 formalism of general relativity. In the 3+1 formalism, the 4-dimensional spacetime is foliated by a family of 3-dimensional spacelike hypersurfaces $\Sigma_{t}$, so that the EFEs can be decomposed into a set of four elliptic-type constraint equations and six evolution equations. \cite{Lin2006}

\section{The field equations and their vacuum solution for the exterior region of the spacetime}
In accordance with definition \eqref{line-element-definition}, the covariant components of the metric tensor that is associated with the line element \eqref{isotropic-line-element} is represented by
\begin{equation} \label{metric-tensor}
	g_{\mu\nu} = \text{diag}(-e^{2\nu},\,e^{2\mu},\, e^{2\mu}r^{2},\, e^{2\mu}r^{2}\sin^2\theta).
\end{equation}
The 9 independent Christoffel symbols, out of the 13 non-zero components of the metric connection, are expressed in the isotropic coordinate frame by
\begin{subequations}
	\begin{align}
		& \displaystyle \Gamma^{t}_{tr} = \nu', \quad \Gamma^{r}_{tt} = e^{2\nu-2\mu}\nu', \Gamma^{r}_{rr} = \mu', \quad \Gamma^{\theta}_{r\theta} = \mu' + \frac{1}{r}, \nonumber  \\[10pt]
		& \displaystyle \Gamma^{r}_{\theta\theta} = -r(r\mu'+1), \quad \Gamma^{r}_{\phi\phi} = -r\sin^2\theta(r\mu'+1), \nonumber  \\[10pt]
		& \displaystyle \Gamma^{\theta}_{\phi\phi} = -\sin\theta \cos\theta, \quad \Gamma^{\phi}_{r\phi} = \mu' + \frac{1}{r}, \quad \Gamma^{\theta}_{\theta\phi} = \cot\theta, \nonumber
	\end{align}
\end{subequations}
where the primes indicate differentiation with respect to $r$. One may, now, continue with the straightforward computation of the covariant components of curvature tensor $R_{\mu\nu}$ (which we do not write down in the interests of brevity) with the purpose of constructing the Einstein field equations 
\begin{equation} \label{Einstein-equations}
	R_{\mu\nu} - \frac{1}{2}Rg_{\mu\nu} = 8\pi T_{\mu\nu},
\end{equation}
where $R$ is the Ricci scalar, $g_{\mu\nu}$, and $T_{\mu\nu}$ are the covariant components of the metric tensor \eqref{metric-tensor}, and the energy--momentum tensor [cf. eqs. \eqref{vacuum-stress-energy-tensor} and \eqref{stress-energy-tensor}], respectively. 

From all the above, one may now determine the non-vanishing mixed-variant components of the EFEs in isotropic coordinates\footnote{These equations, which first appeared in a slightly different form in Eddington's pioneering monograph \cite{Eddington1923}, can be obtained by changing the signature from `Landau–Lifshitz sign convention' to `Pauli convention' and by setting $2\nu \to \nu$ and $2\lambda \to \mu$ in Eddington's equations (43.5).} as
\begin{subequations} \label{gravitational-field-equations}
	\begin{align}
		& \displaystyle 8\pi T^{t}_{t} = -e^{2\nu-2\mu}\frac{2r\mu'' + \mu'(r\mu'+4)}{r},  \\[10pt]
		& \displaystyle 8\pi T^{r}_{r} = \frac{\mu'(r\mu' + 2r\nu' + 2) + 2\nu'}{r},  \\[10pt]
		& \displaystyle 8\pi T^{\theta}_{\theta} = 8\pi T^{\phi}_{\phi} = r\left[r(\mu'' + \nu'') + \mu' + r{\nu'}^{2} + \nu'\right],
	\end{align}
\end{subequations}
where the $(\theta,\, \theta)$-component is identical to the $(\phi,\, \phi)$-component.

\begin{figure}
	\centering
	\includegraphics[scale=0.80]{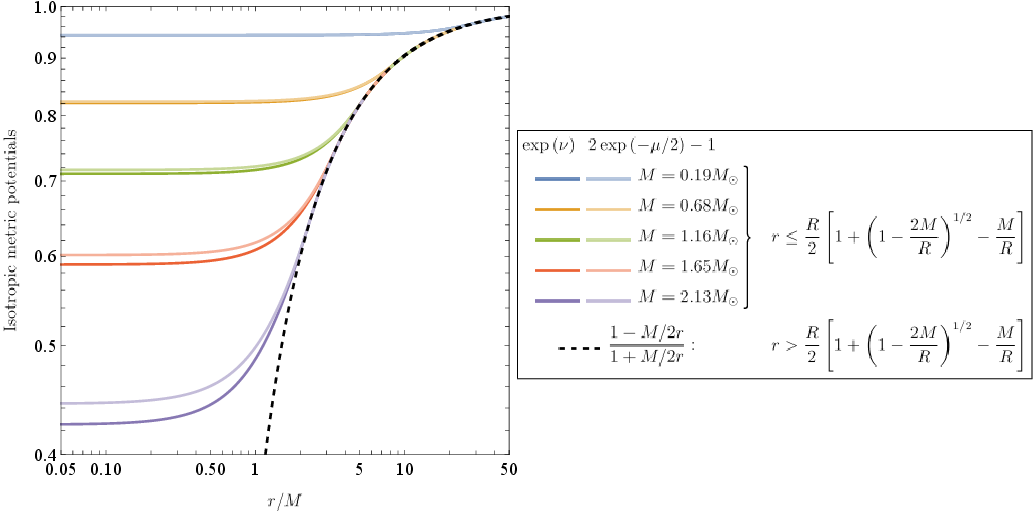}% Here is how to import EPS art
	\caption{The radial dependence of the isotropic metric potentials. Each coloured line correspond to one configuration of relativistic stars, constructed with the SFHo/DD2 EoS for different values of gravitational mass, which were marked by the bullet point symbol $\bullet$ in Figure \ref{fig:mass-radius-relation}. In the interior region of the spacetime, the expressions $\exp(\nu)$ and $\exp(-\mu/2) - 1$ are represented by a curve with a darker and another curve with a lighter shade of the same colour, respectively. At and beyond the stellar surface, located at $r = R$, the Schwarzschild exterior solution \eqref{Schwarzschild-metric-in-isotropic-coordinates} (dashed black line) is valid.}
	\label{fig:isotropic-metrics}
\end{figure}

In the exterior vacuum region ($r > R$), all the mixed-variant components of the energy--momentum tensor are identically zero, that is
\begin{equation} \label{vacuum-stress-energy-tensor}
	T_{\mu\nu} = 0.
\end{equation}
Therefore, the field equations \eqref{gravitational-field-equations} can readily be integrated to yield the Schwarzschild solution \cite{Eddington1923}, expressed in isotropic polar coordinates $(t,\, r,\, \theta,\, \phi)$ by
\begin{equation} \label{Schwarzschild-metric-in-isotropic-coordinates}
	ds^{2} = -\left(\frac{1-M/2r}{1+M/2r}\right)^{2}dt^{2} + \left(1+\frac{M}{2r}\right)^{4}\left[dr^{2} + r^2(\sin^2\theta d\phi^{2} + d\theta^{2})\right],
\end{equation}
where $M$ denotes the total gravitational mass of the massive body. In contrast to its much better-known counterpart, given in curvature coordinates $(\tilde{t},\, \tilde{r},\, \tilde{\theta},\, \tilde{\phi})$ by
\begin{equation} \label{Schwarzschild-metric-in-spherical-coordinates}
	ds^{2} = -\left(1-\frac{2M}{\tilde{r}}\right)dt^{2} + \left(1-\frac{2M}{\tilde{r}}\right)^{-1}d\tilde{r}^{2} + \tilde{r}^2(\sin^2\tilde{\theta} d\tilde{\phi}^{2} + d\tilde{\theta}^{2}),
\end{equation}
the exterior region $r > M/2$ in the isotropic coordinates corresponds to that of $\tilde{r} > 2M$ in curvature coordinates. The limit between the two regions is represented by a black dashed line in Figures \ref{fig:Schwarzschild-metrics}--\ref{fig:isotropic-metrics}. The two systems of coordinates are related to one another via the relations (cf. (B12--B13) in Ref. \cite{Tsokaros2015})
\begin{equation} \label{radius-transformation-exterior}
\tilde{r} = r\left(1+\frac{M}{2r}\right)^{2} \quad \text{and} \quad r = \frac{\tilde{r}}{2}\left[1 + \left(1-\frac{2M}{\tilde{r}}\right)^{1/2} - \frac{M}{\tilde{r}}\right],
\end{equation}
depicted by the dashed black line in Fig. \ref{fig:transformation}, where the rest of the coordinates are identical: $\tilde{t} = t,\, \tilde{\theta} = \theta,\, \tilde{\phi} = \phi$ as it is explained in more detail in \ref{sec:appendix1}. Accordingly, the value of the isotropic radial coordinate $r$ taken at the star's surface is expressed by
\begin{equation} \label{surface-radius}
	R = \frac{\tilde{R}}{2}\left[1 + \left(1 - \frac{2M}{\tilde{R}}\right)^{1/2} - \frac{M}{\tilde{R}}\right]
\end{equation}
if the constants $R$ and $\tilde{R}$ denote the surface radius in curvature and isotropic coordinates, respectively.

\section{The hydrostatic-equilibrium solution for the interior region of the spacetime}
\begin{figure}
	\centering
	\includegraphics[scale=0.80]{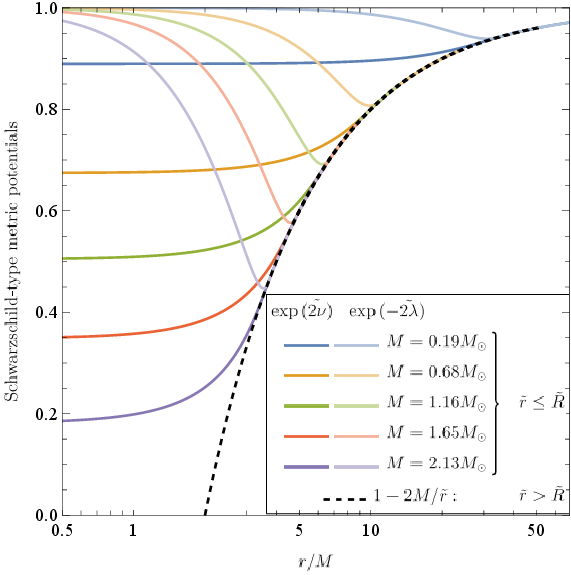}% Here is how to import EPS art
	\caption{The radial dependence of the spherical polar metric potentials. Each coloured line correspond to one configuration of relativistic stars, constructed with the SFHo/DD2 EoS for different values of gravitational mass, which were marked by the bullet point symbol $\bullet$ in Figure \ref{fig:mass-radius-relation}. In the interior region of the spacetime, the expressions $\exp(2\tilde{\nu})$ and $\exp(-2\tilde{\lambda})$ are represented by one curve with a darker, and one with a lighter shade of the same colour, respectively. At and beyond the stellar surface, located at $\tilde{r} = \tilde{R}$, the Schwarzschild exterior solution \eqref{Schwarzschild-metric-in-spherical-coordinates} (dashed black line) is valid.}
	\label{fig:Schwarzschild-metrics}
\end{figure}
In the interior region ($r < R$), let us consider the stress--energy tensor for a perfect fluid in thermodynamic equilibrium, which can be written as
\begin{equation} \label{stress-energy-tensor}
	T_{\mu\nu} = (\varepsilon+p)u_{\mu}u_{\nu} + pg_{\mu\nu},
\end{equation}
with $u^{\mu}$ being the components of the fluid's four-velocity, and the hydrostatic pressure $p$ is related to the mass--energy density $\varepsilon$ through an equation of state. The relations connecting the above quantities of a perfect fluid with the metric of a static, spherically symmetric spacetime \eqref{metric-tensor} are directly obtained by inserting the appropriate components of the stress--energy tensor \eqref{stress-energy-tensor} into the field equations \eqref{gravitational-field-equations}, so that we obtain three independent ODEs as
\begin{subequations} \label{field-eqs}
	\begin{align}
		& \displaystyle 8\pi \varepsilon = -\frac{e^{-2\mu}}{r}\left[2r\mu'' + \mu'(r\mu'+4)\right],  \label{field-eq1}  \\[10pt]
		& \displaystyle 8\pi p = \frac{e^{-2\mu}}{r}\left[\mu'(r\mu' + 2r\nu' + 2) + 2\nu'\right],  \label{field-eq2}  \\[10pt]
		& \displaystyle 8\pi p = \frac{e^{-2\mu}}{r}\left[r(\mu'' + \nu'') + \mu' + r{\nu'}^{2} + \nu'\right].  \label{field-eq3}
	\end{align}
\end{subequations}
The reason for two separate expressions \eqref{field-eq2} and \eqref{field-eq3} connecting the pressure with the metric lies in the fact that the metric tensor \eqref{metric-tensor} is sufficiently general to apply to any static and spherically symmetric distribution, whereas having a perfect fluid involved, implies an equality at each point of the fluid between the stresses in the radial and tangential directions.\cite{Tolman1930a}

Let us first consider expressing the second derivatives of the metric tensor by their first derivatives and by thermodynamic quantities. This can be most conveniently achieved by re-arranging the eq. \eqref{field-eq1} in such a way that yields
\begin{equation} \label{mu-equation}
	\mu'' = -4\pi e^{2\mu}\varepsilon - \frac{2\mu'}{r} - \frac{\mu'^{2}}{2}
\end{equation}
and by substituting this expression into eq. \eqref{field-eq3} in order to obtain
\begin{equation} \label{nu-equation}
	\nu'' = 4\pi e^{2\mu}(\varepsilon + 2p) + \frac{\mu'(r\mu' + 2) + 2\nu'(r\nu' + 1)}{2r}.
\end{equation}
Let us use the last remaining EFE, eq. \eqref{field-eq2}, to express $\nu'$ by
\begin{equation} \label{nu-deriv-equation}
	\nu' = \frac{8\pi e^{2\mu}rp + \mu'(r\mu' + 2)}{2(r\mu' + 1)},
\end{equation}
where $\mu'$ is a functions of $r$ yet to be determined. On account of the fact that eq. \eqref{field-eq2} does not involve higher than first-order derivatives of the metric tensor, an additional equation can be gained by a derivation\footnote{Instead of derivation, one may alternatively compute the equation of hydrostatic equilibrium \eqref{TOV-eq1} from the conservation of the fluid's energy and momentum which demands the covariant divergence of the density- and flux-tensor field to vanish: $\nabla^{\mu}T_{\mu\nu} = 0.$ Taking into account that the spacetime is static implies that $\partial_{t}\varepsilon = \partial_{t}p = 0$, whereas the isotropy implies that $\partial_{\theta}p = \partial_{\psi}p = 0$. Consequently, $\nabla_{\mu}T^{\mu}_{t} = -p' - \frac{1}{2}(p + \varepsilon)\nu' = 0$, as it was first derived by Tolman, Oppenheimer and Volkoff \cite{Tolman1939,Oppenheimer1939} in the spherical polar coordinate frame.} with respect to $r$:
\begin{equation} \label{TOV-eq1}
	p' = -\frac{1}{2}(p + \varepsilon)\frac{8\pi e^{2\mu}rp + \mu'(r\mu' + 2)}{r\mu' + 1},
\end{equation}
where the second derivatives and $\nu'$ were eliminated by the substitution of eqs. (\ref{mu-equation}--\ref{nu-deriv-equation}). 

One might reasonably anticipate finding an expression for $\mu'$ which would involve the gravitational mass $m$ by taking into account that eq. \eqref{mu-equation} poses a direct relationship between the mass--energy density $\varepsilon$ and the metric potential $\mu$ and its derivatives, without involving either $\nu,\, p$ or their respective derivatives. Instead of directly using the corresponding eq. \eqref{field-eq1}, one may as well formally integrate the $(t,\, t)$-component of the EFE, given in curvature coordinates by $8\pi \varepsilon = e^{-2\tilde{\lambda}}\left(\tilde{r}^{-1}\tilde{\lambda}' - \tilde{r}^{-2}\right) + \tilde{r}^{-2}$, with the limiting condition $\tilde{\lambda}|_{\tilde{r}=0} = 0$. The integration yields the expression
\begin{equation} \label{mass-def1}
	e^{-2\tilde{\lambda}} = 1 - \frac{8\pi}{\tilde{r}}\int^{\tilde{r}_{1}}_{0}\varepsilon \tilde{r}^{2} d\tilde{r}
\end{equation}
for $\tilde{\lambda}(\tilde{r}_{1})$ at an arbitrary point $\tilde{r}_{1} \leq \tilde{R}$ inside a spherical star of surface radius $\tilde{R}$. In agreement with the vacuum solution \eqref{Schwarzschild-metric-in-spherical-coordinates}, the metric potential $\tilde{\lambda}$ is connected to the total-gravitational mass of the star $M$ by the relation $e^{-2\tilde{\lambda}} = 1 - 2M/\tilde{R}$, which in turn implies that it is rational to define the total-gravitational mass as
\begin{equation} \label{mass-def2}
	M = \int^{\tilde{R}}_{0}dm = 4\pi\int^{\tilde{R}}_{0}\varepsilon \tilde{r}^{2} d\tilde{r}.
\end{equation}
The differential gravitational mass $dm$ contained in the infinitesimal volume of spherical shell between $r$ and $r+dr$ is given by $dm = 4\pi \varepsilon \tilde{r}^{2} d\tilde{r} = 4\pi \varepsilon (e^{2\mu}r^{2})(e^{\mu}[r\mu'+1]dr)$, in accordance with the coordinate transformation \eqref{radius-transformation-interior}.\footnote{Fig. \ref{fig:Schwarzschild-metrics} illustrates the exponential growth of $\tilde{\lambda}$ by depicting the decrease of the expression $\exp(-2\tilde{\lambda})$ in \eqref{mass-def1} as the gravitational mass \eqref{mass-def2} piles up when one gradually moves from the centre towards the surface. The higher the rate of change, the more compact stellar configuration.} Upon recognising the first bracket as the square of the term $e^{\mu}r$ and the last bracket as the first derivative of the same term, one can considerably simplify the previous expression and obtain
\begin{equation} \label{mass-def3}
	\frac{dm}{dr} = \frac{4\pi\varepsilon}{3}\frac{d}{dr}(e^{\mu}r)^{3}.
\end{equation}
Taking into account that the term $e^{\mu}r$ is equivalent to the areal radius $\tilde{r}$, the expression $d\tilde{A} \equiv d(e^{\mu}r)^{3}/dr$ evidently conveys the meaning of a surface element of the 2-dimensional spatial surface. Consequently, the gravitational-mass element $dm$ in \eqref{mass-def3} preserved the physical interpretation as being the integral functional of the mass--energy density $\varepsilon$ over a volume element $d\tilde{V} \equiv d(e^{\mu}r)^{3}$ in the 3-dimensional space.

Let us emphasise that it is misleading to assume that, in contrast to the relationship \eqref{mass-def1} between $\lambda$ and $\varepsilon$ that is apparent in the context of curvature coordinates, a direct relationship connecting the metric potential $\mu$ and the energy density $\varepsilon$ cannot established for the isotropic coordinates. We demonstrate in \ref{sec:appendix2} that the two quantities are linked together by a polynomial equation of fourth degree in the variable $\exp(2\mu)$, which also involves the radical expressions of $\exp(2\mu)$ and $\epsilon$.

\subsection{Representation of the Tolman–Oppenheimer–Volkoff equation in isotropic coordinates}
Combining eqs. (\ref{mass-def1}--\ref{mass-def2}) with (\ref{lambda-correspondence}) allows us to replace  $\mu'$ by the gravitational mass $m$ in those terms of eqs. (\ref{nu-deriv-equation}--\ref{TOV-eq1}), namely, 
\begin{equation} \label{mu-deriv-equation1}
	\mu'(r\mu' + 2) = \frac{2m}{e^{\mu}r^{2}},
\end{equation}
\begin{equation} \label{mu-deriv-equation2}
	r\mu' + 1 = \left(1 - \frac{2m}{e^{\mu}r}\right)^{1/2},
\end{equation}
which involved first derivatives of the metric-potential function $\mu'$. Eq. \eqref{nu-deriv-equation} is then re-expressed as
\begin{equation} \label{nu-deriv-equation2}
	\nu' = \left(4\pi e^{2\mu}rp + \frac{m}{e^{\mu}r^{2}}\right)\left(1 - \frac{2m}{e^{\mu}r}\right)^{-1/2},
\end{equation}
where $\nu$ acts as an effective relativistic analogue of the Newtonian gravitational potential. Having the derivatives eliminated from eq. \eqref{TOV-eq1} in a similar manner, the condition for hydrostatic equilibrium takes on the form of
\begin{equation} \label{TOV-eq2}
	p' = -\frac{Gm}{e^{\mu}r^{2}}(p + \varepsilon)\left(\frac{4\pi}{c^{4}}\frac{(e^{\mu}r)^{3}p}{m} + 1\right)\left(1 - \frac{2G}{c^{2}}\frac{m}{e^{\mu}r^{2}}\right)^{-1/2},
\end{equation}
with the values of $G$ and $c$ having been restored for comparison's sake. The corresponding Tolman--Oppenheimer--Volkoff equation in curvature coordinates is given by $\displaystyle p' = -\frac{Gm}{\tilde{r}^{2}}(p + \varepsilon)\left(\frac{4\tilde{r}^{3}p}{mc^{2}} + 1\right)\left(1 - \frac{2Gm}{\tilde{r}c^{2}}\right)^{-1}$. By careful examination of eq. \eqref{TOV-eq2} in contrast to its better-known counterpart, the exponent $1/2$ over the last bracket is exposed to be different. Both expressions of the TOV equation are, therefore, of the so-called Riccati-type, a first-order non-linear ODE that is quadratic in isotropic pressure, irrespective of the choice of coordinate frame.

\section{\label{sec:computating-TOV}Numerically computing equilibrium sequences of relativistic stars}
\begin{figure}
	\begin{minipage}{.453\linewidth}
		\centering
		\subfloat[Isotropic pressure against baryon-number density in the crust.]{\label{fig:eos}
			\includegraphics[width=1.00\linewidth]{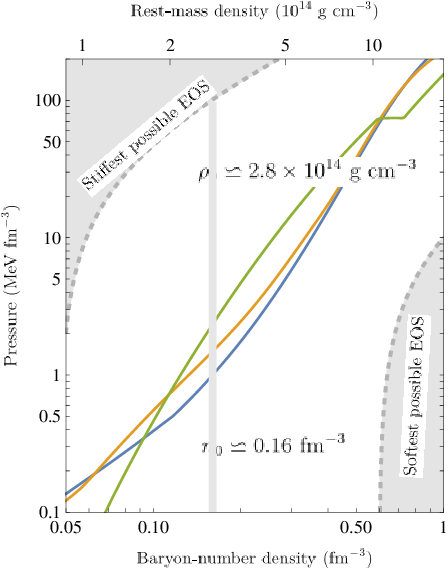}}
	\end{minipage} \hspace*{5pt}
	\begin{minipage}{.50\linewidth}
		\centering  \vspace{20pt}
		\subfloat[Total gravitational mass--central energy density relations.]{\label{fig:mass-radius-relation}
			\includegraphics[width=1.0\linewidth]{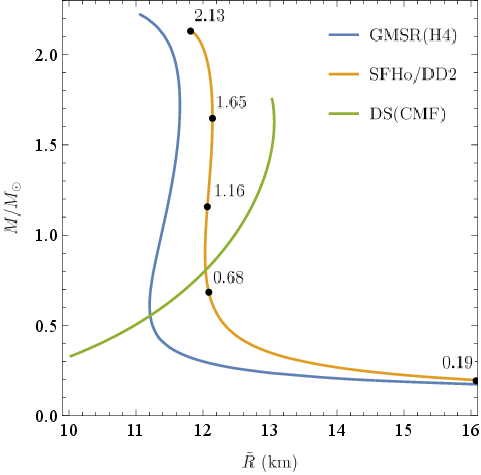}}
	\end{minipage}\par\medskip
	\caption{\textit{Left panel}: The pressure is shown as a function of baryon-number density / rest-mass density for the three EoSs listed in Table \ref{tab:EoS}. The nuclear saturation density $n_{0} \simeq 0.16 \text{ fm}^{-3}$ is denoted by a shaded gray line. The range of pressures at $n_{0}$ is approximately a factor of $1$ to $3$ $\text{MeV fm}^{-3}$. \textit{Right panel}: The corresponding mass--radius relation is displayed for each tabulated EoS with the orange curve representing the sequence of relativistic stars constructed with the EoS SFHo/DD2, whereas the blue and the green curves, respectively, denoting a stiffer EoS (GMSR(H4)) and a softer EoS (DS(CMF)-2 Hybrid) based on different microphysics, both of them are publicly accessible in \href{https://compose.obspm.fr}{\texttt{CompOSE}}. The symbols $\bullet$ mark a set of five selected configurations of relativistic stars that were computed for different values of gravitational mass with the use of the EoS SFHo/DD2; the maximal-mass configuration with $M = 2.13M_{\odot}$, an extremely low-mass configuration with $M = 0.19M_{\odot}$, and three equidistant points in terms of the gravitational mass with $M = 0.68M_{\odot},\, 1.16M_{\odot},\, 1.65M_{\odot}$, respectively. For this set of configurations, the relationship between the two radial coordinates and the behaviour of the metric potentials are shown in Figures \ref{fig:transformation}--\ref{fig:isotropic-metrics}.} \label{fig:eos-and-mass-radius}
\end{figure}
The TOV equation \eqref{TOV-eq2}, together with the expression for the gravitational mass \eqref{mass-def3}\footnote{It should be pointed out that, although the differential eq. \eqref{mass-def3} can be replaced by an algebraic relationship connecting $\mu$ and $\varepsilon$ in the form of the quartic equation \eqref{algebraic-relationship}, the complexity of the algebraic equation makes integrating the differential equation much more practical in terms of computational cost.} and the equations for the metric potentials (\ref{mu-deriv-equation2}--\ref{nu-deriv-equation2}), constitutes a partially uncoupled system of first-order, non-linear and non-autonomous ODEs \cite{Martins2019}, compiled into the following form:
\begin{subequations} \label{TOV-system}
	\begin{align}
		& \displaystyle \frac{dm}{dr} = \frac{4\pi\varepsilon}{3}\frac{d}{dr}(e^{\mu}r)^{3},  \label{TOV-system1}  \\[10pt]
		& \displaystyle \frac{d\mu}{dr} = \frac{1}{r^{2}}\left[\left(1 - \frac{2m}{e^{\mu}r}\right)^{1/2} - 1\right],  \label{TOV-system2}  \\[10pt]
		& \displaystyle \frac{d\nu}{dr} = \left(4\pi e^{2\mu}rp - \frac{m}{e^{\mu}r^{2}}\right)\left(1 - \frac{2m}{e^{\mu}r}\right)^{-1/2},  \label{TOV-system3}  \\[10pt]
		& \displaystyle \frac{dp}{dr} = -\frac{m}{e^{\mu}r^{2}}(p + \varepsilon)\left(\frac{4\pi(e^{\mu}r)^{3}p}{m} + 1\right)\left(1 - \frac{2m}{e^{\mu}r^{2}}\right)^{-1/2}.  \label{TOV-system4}
	\end{align}
\end{subequations}
Supplemented with an EoS that is set out in Table \ref{tab:EoS}, the ODE system \eqref{TOV-system} completely determines the stellar structure in gravitational hydrostatic equilibrium through performing the integration from the centre of the star with initial condition $m(0) = 0$, where an arbitrary value for the central energy-density $\varepsilon_{c} = \varepsilon(0)$, until the pressure $p(r)$ vanishes at some value $r=R$ which corresponds to the position of the star's surface.\footnote{\hspace{0.6pt} Let us remember that surface radius corresponds to different values in curvature and isotropic coordinates. The relation \eqref{surface-radius} links the two values associated with the star's surface.} A unique sequence of relativistic stars in hydrostatic equilibrium corresponds to each EoS, parameterised by $\varepsilon_{c}$. Figure \ref{fig:eos-and-mass-radius} shows the corresponding EoS (left) and mass--radius relations (right), respectively.

With the intention of computing configurations of relativistic stars with considerably large gravitational mass, we used the same EoS (SFHo/DD2) in this paper as in our earlier paper \cite{Kacskovics2023}, where, although isotropic coordinates were adapted in the \href{https://lorene.obspm.fr}{\texttt{LORENE/nrotstar}} code \cite{LUTH1997}, outlining the computational differences stemming from the choice of coordinate frame was not placed at the focus of the study. This zero-temperature EoS is based on a vector-meson extended linear $\sigma$-model that combines two relativistic mean-field (RMF) models, the softer SFHo model developed by \cite{Steiner2013} and the stiffer DD2 model by \cite{Typel2010,Hempel2010}. In an effort to mitigate the risk of drawing inaccurate conclusions about certain numerical features of our computations that may arise from the inherent properties of a single EoS, we also measured the computational time of equilibrium configurations built with a softer and a stiffer EoS, DS(CMF)-2 Hybrid and GMSR(H4), respectively. Both tabulated EoS data are available in the \href{https://compose.obspm.fr}{\texttt{CompOSE}} online repository \cite{CompOSE2022}. DS(CMF)-2 Hybrid \cite{Dexheimer2008} is a relativistic SU(3) chiral mean-field (CMF) model using a non-linear realisation of the $\sigma$-model, which includes pseudo-scalar mesons as the angular parameters for the chiral transformation. GMSR(H4) \cite{Grams2022} corresponds to a unified EoS of cold nuclear matter at $\beta$-equilibrium, where the effective interaction is determined by the non-relativistic meta-model adjusted to reproduce the chiral EFT Hamilatonian H4.
\begin{table}
	\centering
	\begin{tabular}[t]{|@{\extracolsep{\fill}} c @{\extracolsep{\fill}} | c @{\extracolsep{\fill}} c @{\extracolsep{\fill}} c @{\extracolsep{\fill}} c @{\extracolsep{\fill}} c @{\extracolsep{\fill}}|}
		\Xhline{2\arrayrulewidth}
		\rowcolor[gray]{0.96}
		\textbf{ Name } & \textbf{ Particles } &  $\bm{M_{\text{\textbf{max}}}}$ \textbf{ } & $\bm{\min(n_{\text{\textbf{b}}})}$ \textbf{ } & \textbf{ } $\bm{\max(n_{\text{\textbf{b}}})}$ \textbf{ } & \textbf{ Ref. } \\
		\Xhline{2\arrayrulewidth}
		GMSR(H4) & Nucleonic & $2.33 M_{\odot}$ \quad & $10^{-7} \text{ fm}^{-3}$ & 2.00 fm${}^{-3}$ & \cite{Grams2022} \\
		
		SFHo/DD2 & Nucleons & $2.13 M_{\odot}$ & $10^{-9} \text{ fm}^{-3}$ & 1.86 fm${}^{-3}$ & \cite{Steiner2013} \\
		
		DS(CMF)-2 Hybrid & Hybrid & $1.96 M_{\odot}$ & 0.03 $\text{fm}^{-3}$ & 1.6 $\text{fm}^{-3}$ & \cite{Dexheimer2008} \\
		
		\Xhline{2.3\arrayrulewidth}
	\end{tabular}
	% Or to place a caption below a table
	\caption{Some characteristic parameters of the two nucleonic and one hybrid nucleon--hyperon--quark matter models used in the zero-temperature equations of state displayed in Fig. \ref{fig:eos}.} \label{tab:EoS}
\end{table}%

\begin{figure}
	\begin{minipage}{.50\linewidth}
		\centering
		\subfloat[The computation time for evaluating equilibrium configurations, measured as the average CPU execution time.]{\label{fig:tCPU}
			\includegraphics[width=0.95\linewidth]{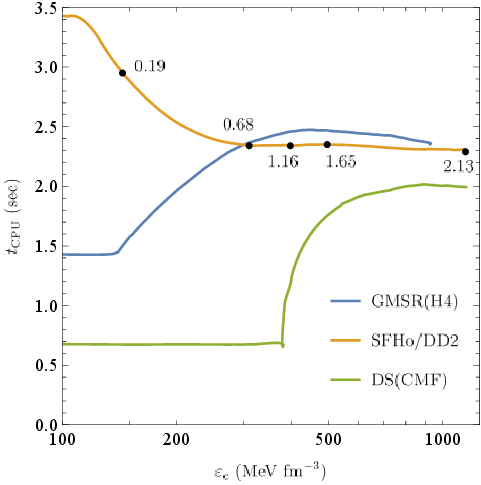}}
	\end{minipage} \hspace*{-2pt}
	\begin{minipage}{.50\linewidth}
		\centering  \vspace{3pt}
		\subfloat[The absolute difference in the gravitational mass between the one computed in the proposed scheme and in \texttt{LORENE}.]{\label{fig:DeltaM}
			\includegraphics[width=1.0\linewidth]{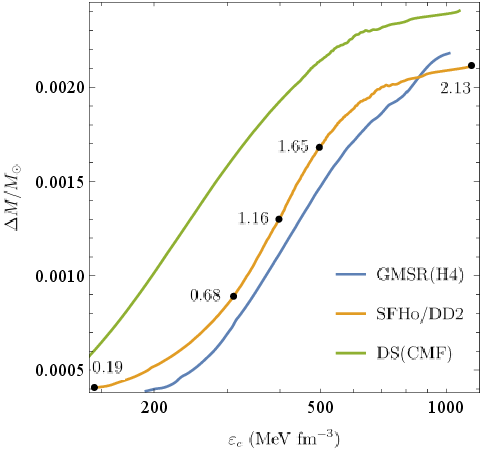}}
	\end{minipage}\par\medskip
	\caption{The performance of computations carried out in the proposed scheme. The computation time required for evaluating equilibrium configurations of relativistic stars for each EoS (left panel) and the absolute difference in the gravitational mass in $M_{\odot}$ units (right panel), are shown as functions of the central energy density $\varepsilon_{c}$. The bullet point symbols $\bullet$ represent the same set of configurations with varying gravitational mass as in Fig. \ref{fig:mass-radius-relation}.} \label{fig:performance}
\end{figure}

The numerical performance of the computational scheme proposed in this paper is qualified in two respects. First, it is qualified by the computation time spent on numerically solving the TOV equations \eqref{TOV-system} for each of the three tabulated EoSs listed in Table \ref{tab:EoS}. Secondly, the accuracy of the numerical results is assessed in terms of the absolute difference in the gravitational mass computed in the proposed scheme compared to the one computed in \texttt{LORENE}. The computation time was measured as the average CPU execution time $t_{\text{CPU}}$, averaged over 1000 runs on an equally-spaced sequence in logarithm scale of the parameter $\varepsilon_{c}$ in the range that lies between $100 \text{ MeV/fm}^{3}$ and the central energy density of the maximum-mass configuration, with respect to the particular EoS. Figure \ref{fig:tCPU} exposes that the length of time required to perform a computation falls in the range of $0.7$ to $3.5$ seconds, where the evaluation, as a general rule, is faster for stiffer EoSs. The computational time for stars with comparatively low or high $\varepsilon_{c}$ stays constant, however, in the intermediate region, the figure features a trend of increase in $t_{\text{CPU}}$ with increasing $\varepsilon_{c}$ for the GMSR(H4) and the DS(CMF)-2, but decrease for SFHo/DD2. The absolute difference in the gravitational mass given by $\Delta M(\varepsilon_{c}) \equiv M_{\text{iso}}(\varepsilon_{c}) - M_{\texttt{LORENE}}(\varepsilon_{c})$ was measured for each EoS in the same range of central energy density as the computation time was. Figure \ref{fig:DeltaM} shows that, following an initial rapid growth in the low-energy-density region, $\Delta M$ increases steadily with $\varepsilon_{c}$ (and therefore with $M$ as well) until the rate of growth once again drops as the central energy density associated with the maximum mass is approached, irrespective of the particular EoS. Another characteristic feature of the $\Delta M$ that is apparent in the figure is that, aside from the $\varepsilon_{c}$, it also increases with stiffness of the EoS.

\section{\label{sec:conclusion}Conclusions}
The results of this study can be divided into an analytical part and a computational part: \\

1. By solving the EFEs set up in isotropic coordinates, we have obtained the equation of hydrostatic equilibrium \eqref{TOV-eq2} which, apart from an exponent and a minus sign, appears unchanged from the familiar form in which the TOV equation appears in terms of curvature coordinates. Complemented by an EoS, the partially uncoupled system of first-order ODEs \eqref{TOV-system} uniquely determines the stellar structure of relativistic stars in hydrostatic equilibrium. We have demonstrated in \ref{sec:appendix2} that the differential equation \eqref{TOV-system1} for the gravitational mass appearing in the ODE system \eqref{TOV-eq2} can be replaced by an algebraic relationship connecting $\mu$ and $\varepsilon$ in the form of the quartic equation \eqref{algebraic-relationship}, but due to the complexity of the algebraic equation, the integration of eq. \eqref{TOV-system1} is much more practical in terms of the computational cost.

In order to contextualise our analytical results, it is worth comparing them with an analysis based on rescaling the radial profile, carried out by Tsokaros, Ury\={u} and Rezzolla \cite{Tsokaros2015}. In order to bring the notations of the quoted article in line with our own, let us change the notations $r,\, \bar{r}$ used for the radial coordinates to our own $\tilde{r},\, r$. Similarly, let us replace the notation $A,\, B,\, \alpha,\, \psi$, used for metric potentials to $\exp(2\tilde{\nu}),\, \exp(2\tilde{\lambda}), \exp(\nu)$ and $\exp(\mu/2)$, respectively. This article demonstrated that the derivation of new structure equations can be avoided by transforming the ODE system of stellar structure from Schwarzschild to isotropic coordinates by rescaling the radial coordinate in accordance with the conversion formula \eqref{rescaling-eq}. Eq. \eqref{rescaling-eq} is the result of eq. \eqref{radius-transformation-interior} and eq. \eqref{mass-def1}, which defines the relation of the Schwarzschild metric potential $\tilde{\lambda}$ to the gravitational mass as $e^{-2\tilde{\lambda}} = 1 - 2m(\tilde{r})/\tilde{r}$. Without the involvement of any relation to the gravitational mass, this transformation would revert back to our relation \eqref{radius-transformation-interior2}.

Ref. \cite{Tsokaros2015} argues that, therefore, one could always carry out the calculations in curvature coordinates and then rescale the radial profile of the solution so that the stellar surface would appear in the correct position. Eq. \eqref{rescaling-eq} allows one to rewrite the structure equations (B8--B11) appearing in \cite{Tsokaros2015} to the form
\begin{subequations} \label{Tsokaros-eqs}
	\begin{align}
		& \displaystyle \frac{dm}{dr} = 4\pi e^{\mu}r^{2}\left(1 - \frac{2m}{e^{\mu}r}\right)^{1/2},  \label{Tsokaros-eq1}  \\[10pt]
		& \displaystyle \frac{d\mu}{dr} = \left[\left(1 - \frac{2m}{e^{\mu}r}\right)^{1/2} - 1\right],  \label{Tsokaros-eq2}  \\[10pt]
		& \displaystyle \frac{d\nu}{dr} = \left(4\pi e^{2\mu}rp - \frac{m}{e^{\mu}r^{2}}\right)\left(1 - \frac{2m}{e^{\mu}r}\right)^{-1/2},  \label{Tsokaros-eq3}  \\[10pt]
		& \displaystyle \frac{dp}{dr} = -\frac{m}{e^{\mu}r}(p + \varepsilon)\left(\frac{4\pi(e^{\mu}r)^{3}p}{m} + 1\right)\left(1 - \frac{2m}{e^{\mu}r}\right)^{-1/2},  \label{Tsokaros-eq4}
	\end{align}
\end{subequations}
that are fully consistent with eqs. \eqref{TOV-system}. Here, eq. \eqref{Tsokaros-eq1} corresponds to the expression \eqref{TOV-system1} and the eqs. \eqref{Tsokaros-eq2}--\eqref{Tsokaros-eq3} were derived with using eq. \eqref{radius-transformation-interior}. \\

2. A further objective of this paper was to evaluate the performance of the computational scheme proposed here in terms of computational time and numerical accuracy. We used the same EoS (SFHo/DD2) as in our earlier paper \cite{Kacskovics2023}, in which paper isotropic coordinates were implemented in the \texttt{LORENE/nrotstar} code, however, it did not consider the impact of different choices of coordinate frame on the computational cost. To avoid unsubstantiated inferences about certain numerical aspects of our computations, which may arise from using a single EoS, two additional EoS models were also included in the study: a softer EoS (DS(CMF)-2 Hybrid) and a stiffer EoS (GMSR(H4)), both of which are available in the \texttt{CompOSE} online repository. Our results indicate that the computational time required to numerically solve the TOV equations tends to increase with the stiffness of the employed EoS. The computational time for stellar models with comparatively low or high $\varepsilon_{c}$ remains constant, but at intermediate values of $\varepsilon_{c}$ the results suggest a tendency for a decrease in the computational time for the SFHo/DD2, whereas an increase for the two other EoS models. Another aspect of evaluating the performance of the present computational scheme was to measure the absolute difference in the gravitational mass between the implementation of this scheme and the results computed by \texttt{LORENE} code for each EoS over the same region of central energy density as the computation time was measured. Our results indicate that after an initial rapid growth in the low energy density region, the absolute difference steadily increases with the central energy density until the rate of growth decreases as the central energy density associated with the maximum mass is reached, irrespective of the particular EoS. Another defining feature of $\Delta M$ is that it not only increases with central energy density, but also with stiffness of the EoS.

\section*{Acknowledgments}
We wish to acknowledge the computational resources provided by the Wigner Scientific Computational Laboratory (WSCLAB) (formerly known as Wigner GPU Laboratory). This research has received partial support from the National Research, Development and Innovation Office (Hungarian abbreviation: NKFIH) under OTKA Grant Agreement No. K138277.

\section*{Data availability statement}
The data underlying this paper will be shared on reasonable request to the corresponding author.

\appendix

\section{Labelling the coordinate frame as ``isotropic'' and its relation to the Schwarzschild-coordinate frame} \label{sec:appendix1}
Provided that the spherical coordinates are converted to Cartesian coordinates by the transformation
\begin{equation} \label{orthogonal-coordinates}
	(r,\, \theta,\, \phi) \mapsto (x^{1},\, x^{2},\, x^{3}): 
	\begin{cases}
		x^{1} = r\sin\theta\cos\phi \\
		x^{2} = r\sin\theta\sin\phi \\
		x^{3} = r\cos\theta,
	\end{cases}
\end{equation}
the Schwarzschild metric \eqref{Schwarzschild-metric-in-isotropic-coordinates}, originally expressed in spherical coordinates, assumes the form of
\begin{equation}
	ds^{2} = -\left(\frac{1-M/2r}{1+M/2r}\right)^{2}dt^{2} + \left(1-\frac{M}{2r}\right)^{4}\left[(x^{1})^{2} + (x^{2})^{2} + (x^{3})^{2}\right],
\end{equation}
where $r = \sqrt{(x^{1})^{2} + (x^{2})^{2} + (x^{3})^{2}}$ (e.g. \cite{Eddington1923}, p. 93). Light-like world lines correspond to the line element $ds^{2} = 0$ and are, therefore, easily expressed in the Cartesian-coordinate frame by
\begin{equation}
	\left(\frac{dx^{1}}{dt}\right)^{2} + \left(\frac{dx^{2}}{dt}\right)^{2} + \left(\frac{dx^{3}}{dt}\right)^{2} = \frac{(1-M/2r)^{2}}{(1+M/2r)^{6}},
\end{equation}
which implies that the coordinates $(t,\, x^{1},\, x^{2},\, x^{3})$ are rightfully called ``isotropic'' on account of the components of the four-velocity field $u^{\mu} = dx^{\mu}/dt$ that are identical in all spatial directions. Evidently, one of the defining features of isotropic line elements is that it exposes uniformity regardless of the orientation in space.

On account that the isotropic space is obtained from the Euclidean 3-space $\mathbb{R}^{3}$ by substituting the usual Euclidean distance with the isotropic distance \cite{Aydin2017}, the metric \eqref{isotropic-line-element} in isotropic coordinates has its spacelike slices conformal to the Euclidean one, where
\begin{equation}
	d\Sigma^{2} = dr^{2} + r^2(\sin^2\theta d\phi^{2} + d\theta^{2})
\end{equation}
is the Euclidean metric. So let us substitute $\tilde{r}$ by $r$ and write down our metric:
\begin{equation} \label{appendix-metric}
	ds^{2} = -\left(1-\frac{2M}{r}\right)dt^{2} + e^{2\mu}[dr^{2} + r^{2}(\sin^2\theta d\phi^{2} + d\theta^{2})].
\end{equation}
By comparing this form of the metric \eqref{appendix-metric} in curvature coordinates with that of \eqref{Schwarzschild-metric-in-spherical-coordinates}, and requiring both the radial and the angular parts to be equal, respectively, we obtain eq. \eqref{radius-transformation-interior}. In order to eliminate the metric function $e^{2\mu}$, the first equation in \eqref{radius-transformation-interior} is divided by the second one, so that
\begin{equation} \label{appendix-metric3}
	\frac{dr^{2}}{r^{2}} = \frac{d\tilde{r}^{2}}{\tilde{r}^{2}-2M\tilde{r}}.
\end{equation}
Finally, taking the square root of eq. \eqref{appendix-metric3} and integrating it will yield the transformation \eqref{radius-transformation-exterior}. Rearranging eq. \eqref{appendix-metric3} and replacing $M$ by $m(\tilde{r})$ yields
\begin{equation} \label{rescaling-eq}
	\frac{d\tilde{r}}{dr} = \frac{\tilde{r}}{r}\sqrt{1-\frac{2m(\tilde{r})}{\tilde{r}}},
\end{equation}
which acts as the relationship between the two radial coordinates inside the star.

\section{Seeking an algebraic relationship between the metric potential $\mu$ and the energy density $\epsilon$} \label{sec:appendix2}
The combination of the eqs. (\ref{mass-def1}--\ref{mass-def2}) with (\ref{lambda-correspondence}) yields an integro-differential equation 
\begin{equation}
	(r\mu' + 1)^{2} = 1 - \frac{8\pi}{e^{\mu}r}\int\varepsilon \cdot e^{3\mu}r^{2}(r\mu'+1)dr,
\end{equation}
then performing the differentiation on both sides results in
\begin{equation} 
	e^{\mu}r^{2}\mu'\left(2r\mu'' + r{\mu'}^{2} + 3\mu'\right) = e^{\mu}(r\mu' + 1) - 8\pi \varepsilon \cdot e^{3\mu}r^{2}(r\mu'+1),
\end{equation}
where the term $\mu''$ can easily be eliminated by eq. \eqref{mu-equation}. As a result, the previous expression turns into an algebraic cubic equation in a single variable $\mu'$, which can be re-arranged into the standard form
\begin{equation} \label{cubic-eq1}
	r^{3}{\mu'}^{3} + 4r^{2}{\mu'}^{2} + r{\mu'} + 1 - 8\pi\varepsilon e^{2\mu}r^{2} = 0.
\end{equation}
Now, restoring the ordinary form of the metric-potential function \eqref{exponential-form} and expressing it, together with its first derivative by
\begin{equation}
	\mu(r) = \frac{1}{2}\log B(r), \quad \mu'(r) = \frac{B'(r)}{2B(r)},
\end{equation}
allows one to eliminate its exponential form from eq. \eqref{cubic-eq1} and re-write it as a general cubic equation
\begin{equation} \label{cubic-eq2}
	a{B'}^{3} + b{B'}^{2} + cB' + d = 0 \quad \text{with} \quad
	\begin{cases}
		a = r^{3}, \\
		b = 8r^{2}B, \\
		c = 4r{B}^{2}, \\
		d = 8B(1 - 8\pi\varepsilon r^{2}B).
	\end{cases}
\end{equation}
After dividing by $a$ and changing the variable $B'$ to
\begin{equation}
	t = B' + \frac{b}{3a},
\end{equation}
eq. \eqref{cubic-eq2} can be transformed into a depressed cubic equation
\begin{equation} \label{depressed-cubic-eq}
	t^{3} + pt + q = 0,
\end{equation}
where the second-degree term is missing and the coefficients of the polynomial are real numbers, given by
\begin{equation} \label{cubic-eq-coefficients}
	p = \frac{3ac - b^{2}}{3a^{2}}, \quad q = \frac{2b^{3} - 9abc + 27a^{2}d}{27a^{3}}.
\end{equation}
Employing Vieta's transformation $t \equiv w - \frac{p}{3w}$ and multiplying by $w^{3}$ transforms the depressed cubic into a quadratic equation
\begin{equation}
	(w^{3})^{2} + q(w^{3}) - \frac{p^{3}}{27} = 0.
\end{equation}
\begin{figure*}
	\begin{minipage}{.453\linewidth}
		\centering
		\subfloat[The discriminant $\Delta$ on log--log scale.]{\label{fig:discriminant}
			\includegraphics[width=1.05\linewidth]{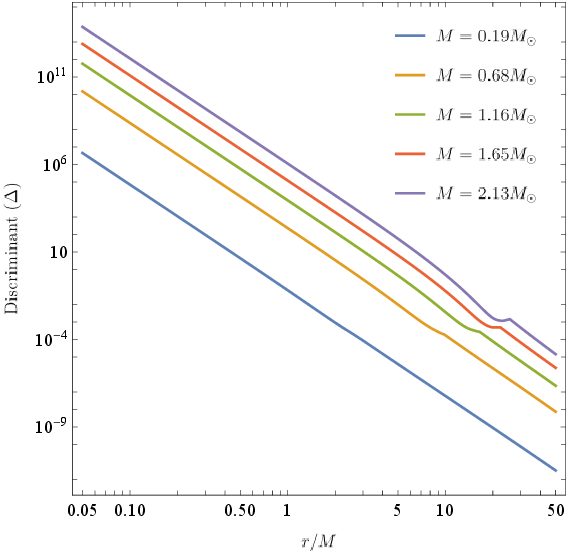}}
	\end{minipage} \hspace*{5pt}
	\begin{minipage}{.50\linewidth}
		\centering  \vspace{10pt}
		\subfloat[The derivative $B'(r)$ as the real root of the cubic equation \eqref{cubic-eq2}.]{\label{fig:Bderiv}
			\includegraphics[width=1.0\linewidth]{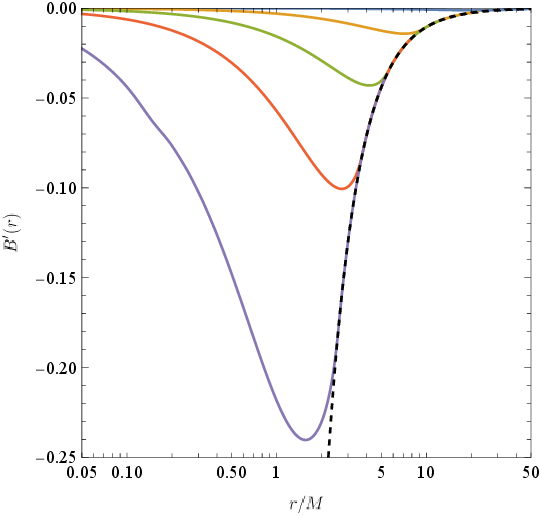}}
	\end{minipage}\par\medskip
	\caption{The discriminant $\Delta$ and the derivative $B'$ are shown on panel (a) and (b), respectively, as functions of $r/M$. Each coloured line corresponds to one configuration of relativistic stars, constructed with the SFHo/DD2 EoS for different values of gravitational mass, which were marked by the bullet point symbol $\bullet$ in Figure \ref{fig:mass-radius-relation}. The dashed black line represents $B'(r \geq R) = (M/r)\left(1+M/2r\right)$ at and beyond the stellar surface, located at $r = R$.
	} \label{fig:cubic-eq}
\end{figure*}
By using a further substitution $u \equiv w^{3}$, the quadratic equation admits two non-zero solutions
\begin{equation} \label{cube-roots}
	u_{\pm} = - \frac{q}{2} \pm \sqrt{\Delta},
\end{equation}
corresponding to the sign of the discriminant $\Delta = q^{2}/4 + p^{3}/27 > 0$. The discriminant, shown on a log--log scale in Figure \ref{fig:discriminant}, appears as an approximately linear decreasing function of radial coordinate $r$. It has a positive intercept, which is greater for stars with larger gravitational mass and a constant, negative slope that does not depend on the mass. Thus, the depressed cubic equation \eqref{depressed-cubic-eq} has one real root and two complex roots. The real root, expressed by Cardano's formula, is the sum of the cube roots \eqref{cube-roots}:
\begin{equation} \label{real-root}
	w_{1} = \sqrt[3]{u_{+}} + \sqrt[3]{u_{-}}.
\end{equation}
Although the conjugate pair of complex roots don't hold any physical relevance, they can be obtained by multiplying the cube root by each of the two primitive cube roots of unity\footnote{\hspace{0.6pt} Let $\delta_{1},\, \delta_{2}$ and $\delta_{3}$ denote the primitive cube roots of unity for $\delta^{3} = 1$. This can be factored into $\delta^{3} - 1 = (\delta - 1)(\delta^{2} + \delta + 1) = 0$, which has the three roots $\delta_{0} = 1,\, \delta_{1} = \frac{-1 + i{\sqrt {3}}}{2}$ and $\delta_{2} = \frac{-1 - i{\sqrt {3}}}{2}$.}:
\begin{equation}
	w_{2} =  -\frac{1}{2}(\sqrt[3]{u_{+}} + \sqrt[3]{u_{-}}) + \frac{i{\sqrt {3}}}{2}(\sqrt[3]{u_{+}} - \sqrt[3]{u_{-}}),
\end{equation}
\begin{equation}
	w_{3} = -\frac{1}{2}(\sqrt[3]{u_{+}} + \sqrt[3]{u_{-}}) + \frac{i{\sqrt {3}}}{2}(-\sqrt[3]{u_{+}} + \sqrt[3]{u_{-}}).
\end{equation}
Reversing the series of transformations implemented above, the real root \eqref{real-root} provides us with an algebraic relationship directly between $\exp(2\mu)$ and $\epsilon$ and, thus, renders the necessity of numerically solving the non-linear ODE \eqref{cubic-eq1} obsolete. This algebraic equation of $\exp(2\mu)$ and $\epsilon$ can be substantially simplified into a form of
\begin{equation} \label{algebraic-relationship}
	\left(\frac{237276}{\beta^{3}} + \frac{6084\sqrt[3]{2}}{\beta} + \beta^{3} + 78\sqrt[3]{4}\beta + 184\right)e^{4\mu} - 432\pi r^{2}\varepsilon e^{2\mu} + 54 = 0,
\end{equation}
which also involves radical expressions of $\exp(2\mu)$ and $\epsilon$ that are incorporated into the terms $\alpha$ and $\beta$:
\begin{subequations}
	\begin{align}
		& \displaystyle \alpha^{3} = -92e^{4\mu} + 216\pi r^{2}\varepsilon e^{2\mu} - 27 - \sqrt{\left(92 e^{4\mu} - 216\pi r^{2}\varepsilon e^{2\mu} + 27\right)^{2} -8788 e^{8\mu}},  \\[10pt]
		& \displaystyle \beta = \alpha + \sqrt[3]{\alpha^{3} - 432\pi r^{2}\varepsilon e^{2\mu}}.
	\end{align}
\end{subequations}

\section*{References} % Produces the bibliography via BibTeX.
\bibliographystyle{iopart-num}
\bibliography{iopart}

\end{document}